\documentclass[twocolumn,aps,prb,showpacs,floatfix,epsfig]{revtex4}

\usepackage{graphicx}%
\usepackage{dcolumn}% Align table columns on decimal point
\usepackage{bm}%

\newcommand \be{\begin{equation}}
\newcommand \ee{\end{equation}}
\newcommand \ba{\begin{eqnarray}}
\newcommand \ea{\end{eqnarray}}
\begin{document}
\title
{\bf Impact ionization fronts in Si diodes:
Numerical evidence of superfast propagation due to nonlocalized preionization}
\author{Pavel Rodin$^{1}$ \cite{EMAIL},
Andrey Minarsky$^{2}$ 
and Igor Grekhov$^{1}$}
%\email{rodin@mail.ioffe.ru}
\affiliation
{$^1$ Ioffe Physicotechnical Institute, Politechnicheskaya 26,
194021, St.-Petersburg, Russia,\\
$^2$ Physico-Technical High School of Russian Academy of Science,
Khlopina 8-3,194021, St.-Petersburg, Russia}

\setcounter{page}{1}
\date{\today}
%\maketitle

\hyphenation{cha-rac-te-ris-tics}
\hyphenation{se-mi-con-duc-tor}
\hyphenation{fluc-tua-tion}
\hyphenation{fi-la-men-ta-tion}
\hyphenation{self--con-sis-tent}
\hyphenation{cor-res-pon-ding}
\hyphenation{con-duc-ti-vi-ti-tes}

%************************************************************************
%----------------------------- ABSTRACT ---------------------------------

\begin{abstract}
We present numerical evidence of a novel propagation mode for superfast
impact ionization fronts in high-voltage Si $p^+$-$n$-$n^+$ structures. In nonlinear dynamics
terms, this mode corresponds to a pulled front propagating into an unstable state 
in the regime of nonlocalized initial conditions. Before the front starts to travel,
field-ehanced emission of electrons from deep-level impurities preionizes
initially depleted $n$ base creating spatially nonuniform free carriers profile.
Impact ionization takes place in the whole high-field region.
We find two ionizing fronts that propagate in opposite directions with
velocities up to 10 times higher than the saturated drift velocity. \end{abstract}
%\endabstract


%72.20.Ht- High Field and Nonlinear Effects
%85.30.-z    Semiconductor devices

\maketitle

Superfast impact ionization fronts travel with velocity $v_f$ higher 
than  the saturated drift velocity $v_s$. Excitation of such fronts
is the fastest non-optical method to modulate conductivity of a high-voltage 
semiconductor structure. \cite{LEV05,Si,GaAs}
This method has important pulse-power application.\cite{GRE89,FOC97,Kardo}
Recently, we proposed a novel mode of superfast front propagation in semiconductor
structures: a pulled
front propagating into unstable state in the regime of nonlocalized
initial conditions. \cite{ROD08}
This mode is expected to appear in $p^{+}$-$n$-$n^{+}$ structures with a relatively low
$n$ base doping level. The propagation mechanism is qualitatively different from
that for the well-known TRAPATT-like propagation mode, 
\cite{DEL70,KYU07,ROD07}
widely considered as a most feasible mechanism of the superfast switching of high-voltage
pulse-power devices. \cite{LEV05}
For the impact ionization front the term ``nonlocalized initial conditions" \cite{SAA03} means
a small free carriers concentration that decreases in the direction of front propagation.
It has been suggested \cite{ROD08} that such preionization of the initially depleted
$n$ base may be created by field-enhanced electron emission from deep-level centers.\cite{ROD05}
Numerical simulations presented in the Letter confirms this suggestion
and provide the first evidence of superfast front propagation due to nonlocalized preionization
in realistic Si $p^{+}$-$n$-$n^{+}$ structures used in pulse-power applications.

\begin{figure}
\begin{center}
\includegraphics[width=3.0 cm,height=5.0 cm,angle=270]{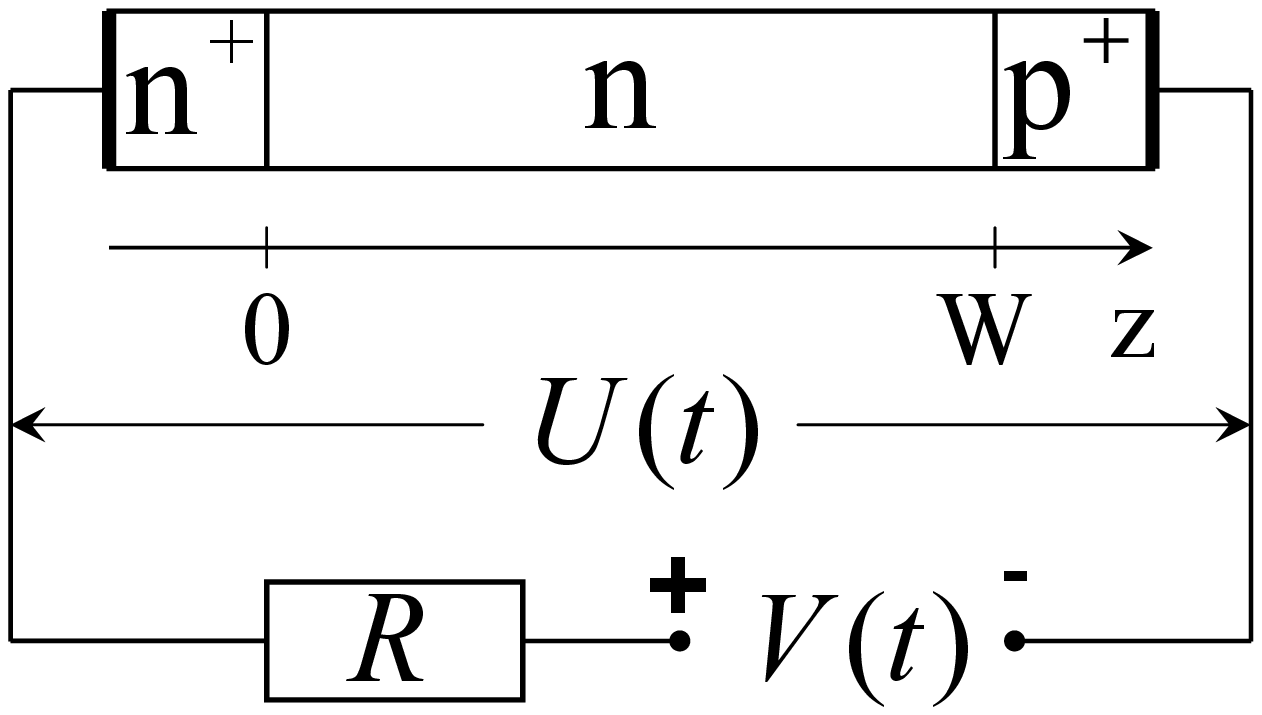}
\end{center}
\caption{Sketch of the reversely biased $p^+$-$n$-$n^+$-structure
operated in the external circuit.}
\end{figure}

Process-induced (PI) deep-level centers in Si are
double-charged donors with ionization energies
0.28~eV (U level) and 0.54~eV (M level).\cite{SAH,sulfur} 
Originally thought of as recombination centers, \cite{SAH}
PI centers turned out to be deep electron traps. \cite{sulfur} As they
do not influence the life-time of nonequilibrium carriers (the most carefully controlled
parameter of the commercial material), their presence in high-purity
Si is ``hidden''.\cite{sulfur} In high voltage structures used in power
applictations PI centers appear in concentration $N_{\rm PI}=10^{11}...10^{13} \; {\rm cm}^{-3}$
(Ref.~\onlinecite{sulfur}). 
Field-enhanced ionization of PI centers is a potential
mechanism of  determenistic low-jitter triggering of the ionizing front
in high-voltage structures. \cite{ROD05}
In the simplest case of low temperatures ($T < 200$~K) 
the emission rate depends on the electrical field $F$ as(Ref.~\onlinecite{ROD05})
\begin{equation}
\label{ionizationRate}
e(F)=\frac{F}{\sqrt{8 m E_0}}\exp \left(-\frac{F_0}{F}\right)
\exp\left(2 \sqrt{\frac{E_{\rm B}}{E_0}}\ln{\frac{6F_0}{F}}\right),
\end{equation}
where 
$F_0 \equiv 4 \sqrt{2 m E^3_0}/3 q \hbar$,
$E_0=0.28$~eV is the binding energy (U level), $E_{\rm B}$ is Bohr energy in semiconductor,
$m$ is the effective electron mass and $q$ is the elementary charge.
Field-enhanced ionization of PI centers has a characteristic threshold
$F_{\rm th}^{\rm PI} \approx 3 \cdot 10^5 \; {\rm V/cm}$ (Ref.\onlinecite{ROD05}) that exceeds
the threshold of band-to-band impact ionization 
$F_{\rm th}^{\rm imp} \approx 2 \cdot 10^5 \; {\rm V/cm}$.
It is important that
$F_{\rm th}^{\rm PI} > F_{\rm th}^{\rm imp}$ because 
electrical field in the $n$ base should be increased above $F_{\rm th}^{\rm imp}$
before impact ionization starts.\cite{ROD05} 

\begin{figure}
\begin{center}
\includegraphics[width=7.5 cm,height=3.75 cm,angle=0]{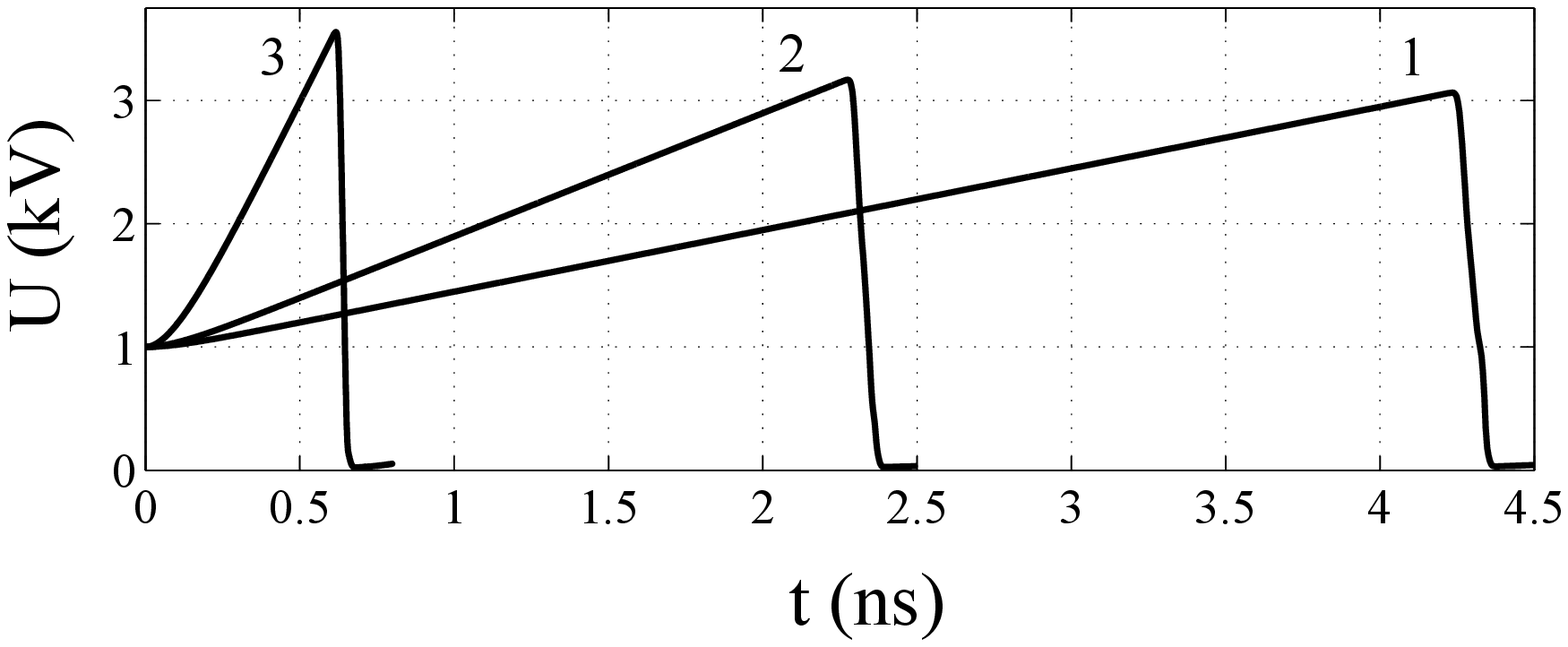}
\end{center}
\caption{Voltage $U(t)$ during the switching process
for different voltage ramps $A=0.5,~1,~5$~kV/ns (curves 1,2 and 3, respectively).}
\end{figure}

Superfast triggering occurs when a sharp voltage ramp is applied
to reversely biased Si $p^{+}$-$n$-$n^{+}$ structure
connected to the voltage source $V(t)$ in series with load resistance $R$ (Fig.~1).
The applied voltage is modelled as
$V(t)=V_0+A \cdot t$,  where the inital voltage $V_0$ is chosen so that the $n$ base
is fully depleted at $t=0$.
We assume that $p^+$-$n$ and $n$-$n^{+}$ junctions are sharp and restict modelling to the $n$ base.
To describe the dynamics of electron and hole concentrations $n(z,t)$,
$p(z,t)$ and electrical field $F(z,t)$, we use the standard one-dimensional
drift-diffusion model together with the Poisson equation and the Kirchoff equation
for the external circuit. The generation term
includes the cut-off that eliminates unphysical multiplication at low
concentrations $n,p < n_{\rm cut}$.\cite{ROD02} 
We refer to Ref.~\onlinecite{ROD02} for the details of the model
and of the numerical method used. The generation term additionally incorporates
field-enhanced emission of free electrons from PI centers with a rate
given by Eq.~(\ref{ionizationRate}). We use the following set of structure
and circuit parameters:
the $n$ base length $W=100 \, {\rm \mu m}$, dopant concentration
$N_{\rm d}=10^{13} \, {\rm cm}^{-3}$,
$N_{\rm PI}=10^{12} \, {\rm cm}^{-3}$, cross-section area 
$S=0.02 \, {\rm cm}^{-3}$, load resistance $R=50$~Ohm, 
initial bias $V_0=1$~kV. The respective stationary breakdown voltage is about 1.2~kV.
We choose $n_{\rm cut}=10^9~{\rm cm^{-1}}$, as in Refs. \onlinecite{ROD02,ROD02a}.

In Fig.~2 we show the voltage $U(t)$ over the diode during switching process.
At the first stage $U(t)$ nearly follows the applied voltage
$V(t)$. Then, $U(t)$ sharply drops and the current flowing through the diode
increases. The triggering time is close to 100~ps and decreases with increase
of the voltage ramp $A$. This is 10 times faster than the drift time $W/v_s \approx 1$~ns. 
Such superfast switching has been observed in the whole range of actual voltage ramps 0.5...10 kV/ns.
Although the transient $U(t)$ looks similar
to the earlier numerical results for TRAPATT-like \cite{BIL83,KAR96,FOC97,GAU98,ROD02} or
tunneling-assisted \cite{ROD02a,RUK05} impact ionization fronts,
the inner dynimics turns out to be qualitatively different.

\begin{figure}
\begin{center}
\includegraphics[width=8.5 cm,height=7.5 cm,angle=0]{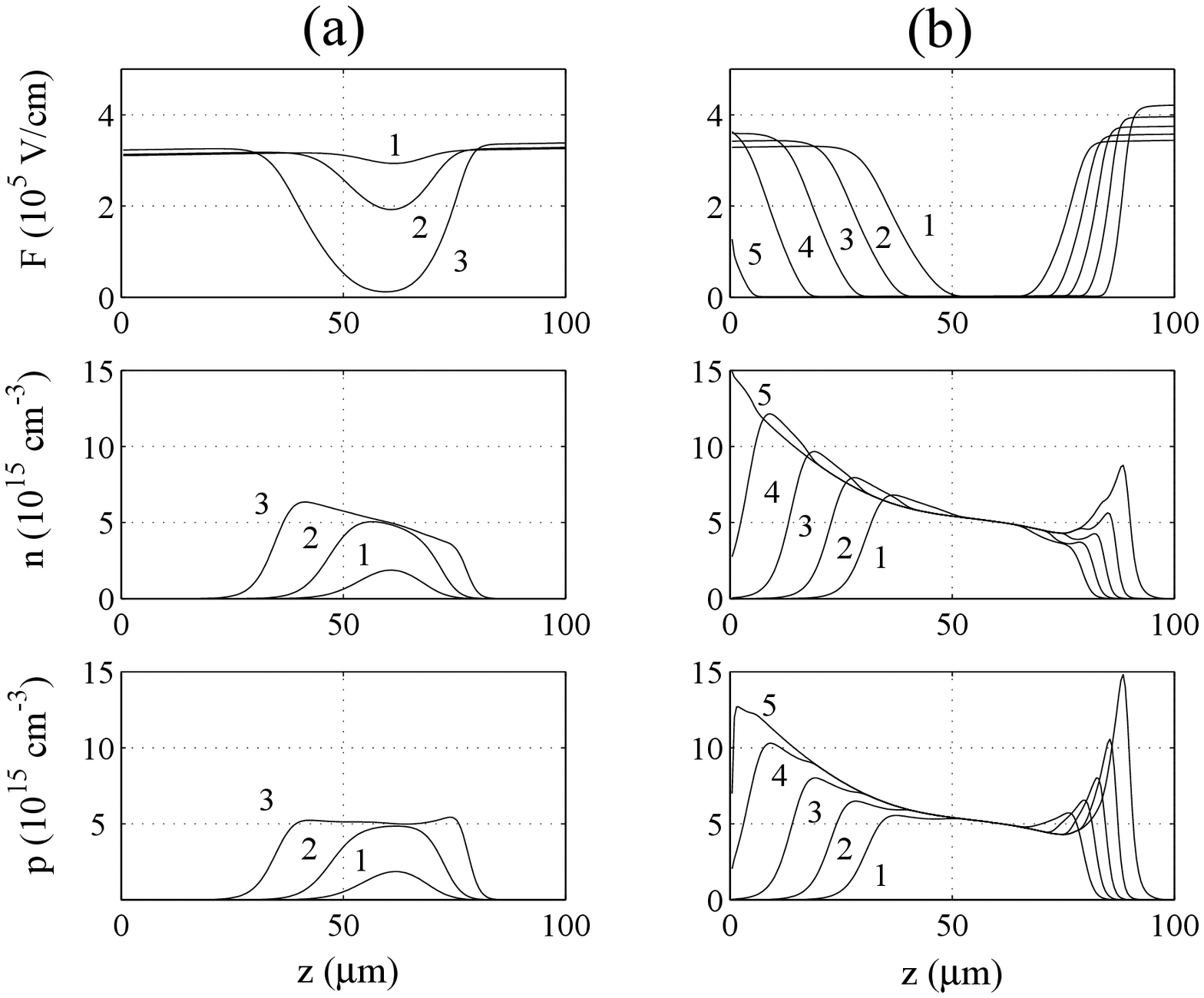}
\end{center}
\caption{The spatial profiles of the electrical field $F(z,t)$ and 
electron and hole concentrations $n(z,t)$ and $p(z,t)$ in the 
$n$-base at different times: (a) increase of the free carrier concentration
due to the field-enhanced ionization of deep-level impurities and subsequent
avalanche multiplication
at times $t=2.285,2.295,2.31$~ns (curves 1,2,and 3, respectively);
(b) propagation of impact ionization fronts
at $t=2.315,2.325,2.335,2.345,2.355$~ns (curves 1,2,3,4, and 5, repectively).
Numerical parameters as in Fig.~2 for $A=1$~kV/ns.}
\end{figure}

In Fig.~3 we show the inner dynamics for $A=1$~kV/ns. 
Due to the higher electrical field, the emission of electrons
from PI centers is most efficient near $p^{+}$-$n$ junction. Free electrons drift
to the  left and multiplicate by the band-to-band impact ionization.
Due to this drift, the maximum of the concentration profile is shifted from the 
$p^{+}$-$n$ junction  into the $n$ base [Fig.~3(a)].
Screening begins when concentrations $n$ and $p$ overcome $N_{\rm d}$.
Eventually, the avalance multiplication creates the initial plasma layer
that fully screens the applied electrical field [Fig.~3(a), curve 3].
Afterwards, the plasma layer expands due to propagation of two
ionizing fronts travelling in opposite directions [Fig.~3(b)]. 

\begin{figure}
\begin{center}
\includegraphics[width=7.5 cm,height=3.75 cm,angle=0]{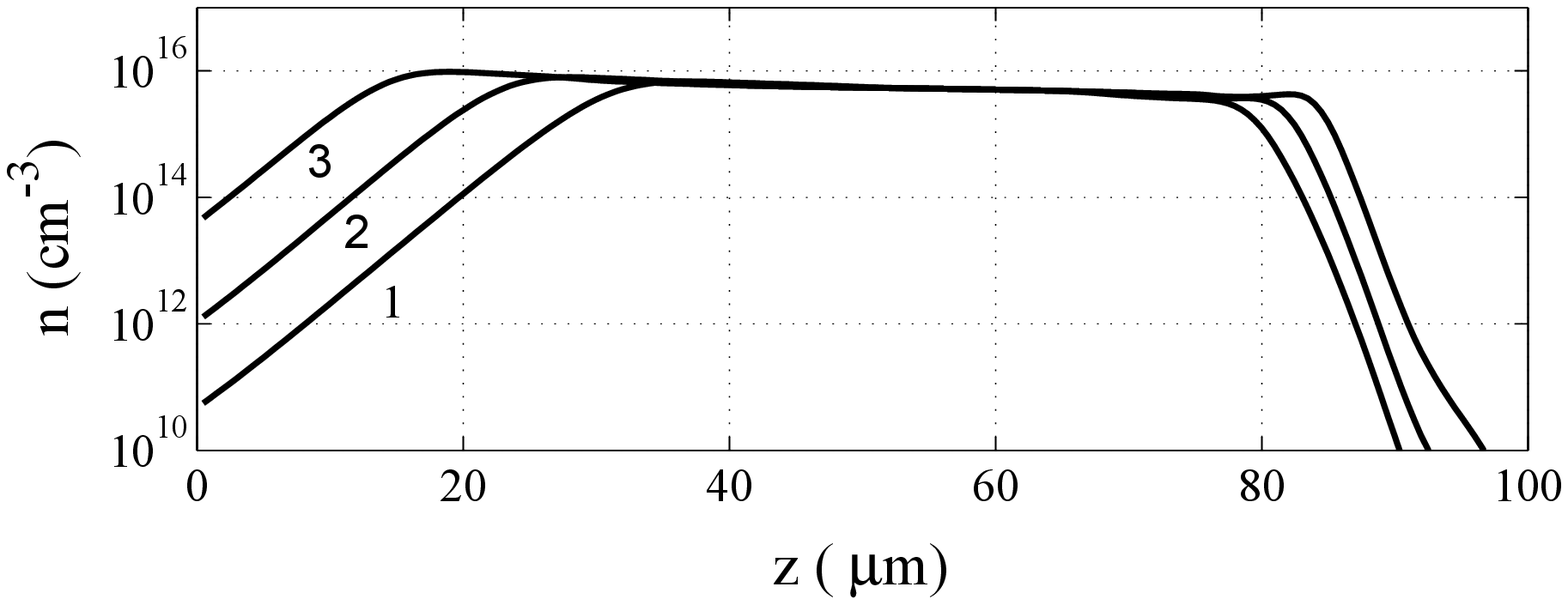}
\end{center}
\caption{The spatial profile of 
electron concentrations $n(z,t)$ in the propagating front 
at times $t=2.315,2.325,2.335$~ns (curves 1,2,and 3, respectively,)
in logarithmic scale. Numerical parameters as in Fig.~3.}
\end{figure}

The left and the right (negative and positive) fronts travel
with velocities $v^{-}_f \approx 10 \, v_s$ and $v^{+}_f \approx 3 \, v_s$, respectively.
Both fronts propagate into the areas where the electrical field $F_{\rm m}$
is nearly constant in space and varies in the interval from 
$3\cdot 10^5$ to $4\cdot 10^{5}$~V/cm,
nearly 2 times larger than $F_{\rm th}^{\rm imp}$.
The small amount of free cariers is present 
everywhere (Fig.~4). Hence avalance multiplication also goes on
all over the high-field region. These conditions are favorable for 
quasiuniform breakdown. The reason why the front-like solution nevertheless
occurs is the nonuniform profile of preionization: the concentration of free carriers in the high-field region
decreases in the direction of front
propagation. This decay happens to be nearly exponential $n,p \sim \exp(\pm \lambda z)$ with
characteristic exponents $\lambda^{-} \approx 4 \cdot 10^3~{\rm cm^{-1}}$ and 
$\lambda^{+} \approx 1.3 \cdot 10^4~{\rm cm^{-1}}$ for negative and positive fronts, respectively (Fig.~4). 
In nonlinear dynamics, such fronts are known as pulled fronts propagating into unstable
state in the regime of nonlocalized initial conditions. \cite{SAA03} 

Let us compare the numerical results with analytical predictions for 
front velocity and plasma concentration made in Ref.~\onlinecite{ROD08}
for the simplified model where electrons and holes
have the same impact ionization coefficients $\alpha(F)=\alpha_{\rm n}(F)=\alpha_{\rm p}(F)$ and
$N_d=0$.
For sufficiently small exponent $\lambda$, the front velocity is given by
\begin{equation}
\label{velocity}
{v_f}/{v_s} \approx {2 \alpha_{\rm m}}/{\lambda},
\end{equation}
where $\alpha_{\rm m} \equiv \alpha(F_{\rm m})$ is the impact ionization coefficient
in the high field region the front propagates to. In Si in strong electrical
fields impact ionization by holes is much weaker than by electrons. 
Therefore  $2 \alpha_{\rm m}$
shall be replaced by $\alpha_{\rm m} \equiv \alpha_{\rm n}(F_{\rm m})$
in the nominator of Eq.~(\ref{velocity}).
The numerical value is $\alpha_{\rm m} \approx 3.2 \cdot 10^4~{\rm cm^{-1}}$ 
for $F_{\rm m} = 3.5 \cdot 10^5~{\rm V/cm}$. 
Eq.~(\ref{velocity}) is applicable if $\lambda \ll \lambda^{\star}$,
where $\lambda^{\star}$ corresponds to so called marginally stable front \cite{SAA03}
and is explicitly determined by $\alpha_{\rm m}$,
diffusion coefficient $D$ and $v_s$ (see Ref. \onlinecite{ROD08}). 
Simple calculation shows that $\lambda^{\star} \approx 2 \cdot 10^5~{\rm cm^{-3}}$.
Hence, the condition $\lambda^{\pm} \ll \lambda^{\star}$ is met for the case under 
consideration, and we obtain analytical estimates
$v_f^{-} \approx \alpha_{\rm m}/\lambda^{-} \approx 8 \, v_s$
and $v_f^{+} \approx \alpha_{\rm m}/\lambda^{+} \approx 2.5 \, v_s$ for the negative and positive
fronts shown in Figs.~3,4, respectively,
in good agreement with numerical results. 

For the symmetrical model, the plasma concentration $\sigma \equiv n+p$ behind
the front is given by 
$
\sigma_{\rm pl}= (2 \varepsilon \varepsilon_0 \alpha_0 F_0 / q) \int^{F_{\rm m}/F_0}_{0} \exp(-1/x) dx,
$
where the ionization coefficient is approximeted as 
$\alpha(F)=\alpha_0 \exp(-F_0/F)$ (Ref.~\onlinecite{ROD08}).
Removing again the factor of 2 and taking the numerical values $\alpha_0= 7.4 \cdot 10^{5}~{\rm cm}$,
$F_0=1.1 \cdot 10^{6} $~V/cm for electrons in Si, 
we get $\sigma_{\rm pl} \approx 10^{16}~{\rm cm^{-3}}$ for the actual electrical field $F_{\rm m} = 3.5 \cdot 10^5$~V/cm, 
in good agreement with numerical results. 

The velocity of the convential TRAPATT-like front is determined by the size of impact
ionization zone. \cite{DEL70,KYU07,ROD07} In contrast, the velocity of the pulled front
in the regime of nonlocalized initial conditions is determined by the slope of the preionization profile,
while the size of impact ionization zone coincides with the total size of the high-field region.
Consequently, the pulled fronts  shall be expected in structures with moderate
doping $N_d$ where the condition $F>F_{\rm th}^{\rm imp}$ is met all over
the high-field region. However, a certain level of the $n$ base
doping $N_d$ is crucial because it leads to the slope of the electrical field in the depleted $n$ base.
Due to this slope the field-enhanced ionization of PI centers is spatially
nonuniform and creates decaying profile of preionization. In particluar, in $p$-$i$-$n$ structures
nonlocalized initial conditons for the pulled front propagation cannot be created by deep-level impurities.
For large $N_d$ the convential TRAPATT-like mode \cite{DEL70,KYU07,ROD07}
occurs, but the cross-over to the pulled mode is possible at the last stage of front propagation
when the condition  $F>F_{\rm th}^{\rm imp}$ is met everywhere.

We  are indebted  to W.~Hundsdorfer who has developed
numerical tools used in the simulations. This work was supported
by Russian Academy of Science in the framework of the project ``Power semiconductor
electronics and pulse technology".
P.~Rodin is grateful to A.~Alekseev for hospitality at the University of Geneva
and acknowledges the support from the Swiss National Science Foundation.


\begin{thebibliography}{99}

\bibitem[*]{EMAIL}
Electronic address: {\rm rodin@mail.ioffe.ru}

\bibitem{LEV05}
M.~Levinshtein, J.~Kostamovaara, S.~Vainshtein, {\it Breakdown
Phenomena in Semiconductors and Semiconductor Devices} (Word
Scientific, 2005).

\bibitem{Si}
I.V.~Grekhov and A.F.~Kardo-Sysoev, Sov.~Tech.~Phys.~Lett. {\bf
5}, 395 (1979).

\bibitem{GaAs}
Zh.I.~Alferov, I.V.~Grekhov, V.M.~Efanov, A.F.~Kardo-Sysoev,
V.I.~Korol'kov, and M.N.~Stepanova
Sov.~Tech.~Phys.~Lett. {\bf 13}, 454 (1987).

\bibitem{GRE89}
I.V.~Grekhov, Solid-State Electron. {\bf 32}, 923 (1989).

\bibitem{FOC97}
R.J.~Focia, E.~Schamiloghu, C.B.~Fledermann, F.J.~Agee and
J.~Gaudet, 
IEEE Trans.~Plasma~Sci. {\bf 25}, 138 (1997).


\bibitem{Kardo}
A.F.~Kardo-Susoev, {\it New Power Semiconductor Devices for 
Generation of Nanosecond Pulses}, in {\it Ultra-Wideband Radar
Technology}, edited by James D. Taylor, CRC Press, Boca Raton,
London, New York, Washington, 2001, pp.205-209.

\bibitem{ROD08} P.~Rodin, A.~Minarsky, I.~Grekhov,
Appl.~ Phys.~Lett. {\bf 93}, 013503 (2008). 

\bibitem{ROD05} P.~Rodin and I.~Grekhov, Appl.~Phys.~Lett. {\bf 86},
243504 (2005);
P.~Rodin, A.~Rodina, and I.~Grekhov, J.~Appl.~Phys. {\bf 98},
094506 (2005).

\bibitem{DEL70}
B.C.~Deloach and D.L.~Scharfetter, IEEE Trans.~Electron Devices
{\bf ED-20}, 9 (1970).

\bibitem{KYU07}
A.S.~Kyuregyan, Semiconductors {\bf 41}, 737 (2007).

\bibitem{ROD07}
P.~Rodin, U.~Ebert, A.~Minarsky and I.~Grekhov, J.~Appl.~Phys.
{\bf 102}, 034508 (2007).

\bibitem{SAA03}
W.~van~Saarloos, Physics Reports {\bf 386}, 29 (2003)

\bibitem{SAH}
L.D.~Yau and C.T.~Sah,  Solid-State Electronics {\bf 17}, 193(1974);
J.~Apl.~Phys. {\bf 46}, 1767 (1975).

\bibitem{sulfur}
E.V.~Astrova, V.B.~Voronkov, V.A.~Kozlov and A.A.~Lebedev,
Semicond.~Sci.~Technol. {\bf 13}, 488-495 (1998).

\bibitem{BIL83}
Yu.D. Bilenko, M.E. Levinstein, M.V. Popova and V.S.
Yuferev, Sov.~Phys.~Semicond. {\bf 17}, 1156 (1983).

\bibitem{KAR96}
A.F. Kardo-Susoev and M.V. Popova,
Sov.~Phys.~Semicond. {\bf 30}, 431 (1996).

\bibitem{GAU98}
H.~Jalali, R.~Joshi, and J.~Gaudet,
IEEE~Trans.~Electron Devices {\bf 45}, 1761-1768(1998).

\bibitem{ROD02}
P.~Rodin, U.~Ebert, W.~Hundsdorfer, and I.~Grekhov,
J.~Appl.~Phys. {\bf 92}, 1971 (2002).

\bibitem{ROD02a}
P.~Rodin, U.~Ebert, W.~Hundsdorfer, and I.~Grekhov,
J.~Appl.~Phys. {\bf 92}, 958 (2002).

\bibitem{RUK05}
S.K.~Lyubutin, S.N.~Rukin, B.G.~Slovikovsky,
S.N.~Tsyranov, Tech.~Phys.~Lett. {\bf 31}, 196 (2005).  

\end{thebibliography}
\end{document}